\documentclass[preprint,structabstract]{aa}
\usepackage{txfonts}
\usepackage{units}
\usepackage{graphicx}
\usepackage[unicode=true]{hyperref}
\usepackage{natbib}
\bibpunct{(}{)}{;}{a}{}{,}
\begin{document}

\title{Progenitor delay-time distribution of short gamma-ray bursts: Constraints from
observations}

\subtitle{}

\author{Jing-Meng Hao\and Ye-Fei Yuan}

\institute{Key Laboratory for Research in Galaxies and Cosmology CAS, \\
Department of Astronomy, University of Science and Technology of China,\\
 Hefei, Anhui 230026, China \\
\email{yfyuan@ustc.edu.cn}}

\offprints{Ye-Fei Yuan}

\date{Received ...; Accepted...}

\abstract{The progenitors of short gamma-ray bursts (SGRBs) have not yet been well
identified. The most popular model is the merger
of compact object binaries (NS-NS/NS-BH). However, other progenitor models cannot be ruled out.
The delay-time distribution of SGRB progenitors, which is an important property to constrain
progenitor models,
is still poorly understood.} {We aim to better constrain the
luminosity function of SGRBs and the
delay-time distribution of their progenitors with newly discovered SGRBs.}
{We present a low-contamination sample of 16 \emph{Swift} SGRBs that
is better defined by a duration shorter than 0.8~s. By using this robust sample and by
combining a self-consistent star formation model with various models for the distribution of time delays,
the redshift distribution of SGRBs is calculated and then compared to the observational data.} {
We find that the power-law delay distribution model is disfavored and that only the lognormal delay
distribution model with the typical delay $\tau\gtrsim\unit[3]{Gyr}$ is consistent with the data.
Comparing \emph{Swift} SGRBs with $T_{90}>0.8\,\mathrm{s}$ to our robust sample ($T_{90}<0.8\,\mathrm{s}$),
we find a significant difference in the time delays between these two samples.}
{Our results show that the progenitors of SGRBs are dominated by relatively long-lived systems ($\tau\gtrsim\unit[3]{Gyr}$),
which contrasts the results found for Type Ia supernovae.
We therefore conclude that primordial NS-NS systems are not favored as the dominant SGRB progenitors. Alternatively,
dynamically formed NS-NS/BH and primordial NS-BH systems with average delays longer than $\unit[5]{Gyr}$ may
contribute a significant fraction to the overall SGRB progenitors.}


\titlerunning{Progenitor delay-time distribution of short GRBs}
\authorrunning{Hao \& Yuan}
\maketitle

\section{Introduction}

Gamma-ray bursts (GRBs) are the most energetic explosions in the Universe,
which can be divided into two major classes: short-duration ($<\unit[2]{s}$)
bursts having a harder spectrum and long-duration ($\geq\unit[2]{s}$)
bursts having a softer spectrum
\citep[e.g.][]{2007PhR...442..166N}.
The bimodality of GRB duration suggests that the progenitors of these
two classes are
likely to be distinct. Long GRBs (LGRBs) occurring in actively star-forming
galaxies with high redshift \citep[e.g.][]{2002MNRAS.334..983T} and
their association with core-collapse supernovae \citep{2003Natur.423..847H,2006ARA&A..44..507W}
suggest a strong connection between them and the collapse of massive
stars \citep[collapsars;][]{1999ApJ...524..262M}, and LGRB rate is hence expected
to trace the star formation rate. On the other hand, short GRBs (SGRBs)
were found in elliptical galaxies \citep{2005Natur.438..988B} with very
low star formation rates, demonstrating that at least some of their progenitors
belong to an old stellar population ($\gtrsim\unit[1]{Gyr}$), and
hence a time delay between the occurrence of the short burst and the epoch of
star formation activity in their hosts is expected, implying that
the progenitors of SGRBs are different from those of LGRBs.
The most popular model for SGRBs is the
merger of either double neutron star (NS-NS) or neutron star-black hole
(NS-BH) binaries \citep{1992ApJ...395L..83N}.
However, other possible progenitor models exist, which include accretion-induced collapse
(AIC) of neutron stars \citep{1998ApJ...494L..57Q}, magnetars and
quark stars \citep[e.g.][and references therein]{2007PhR...442..166N}.

The delay-time distribution of SGRB progenitors has not yet been well understood,
theoretically and observationally.
For the model of the merger of compact object binaries,
the time delay between the formation
of the two main-sequence stars and the merger of the two evolved compact objects
is driven by the emission of the gravitational
wave (GW), which is strongly dependent on the initial separation of the binary.
A $\tau^{-1}$ type of delay-time distribution is a general prediction for this kind of source,
as suggested by recent studies on the rate of type Ia supernovae, where the delay
time is also determined by the gravitational radiation of binaries \citep{2008PASJ...60.1327T}.
Both population synthesis models \citep{2006ApJ...648.1110B} and
observations of six NS-NS binaries in the Milk Way \citep{2004MNRAS.350L..61C}
support this type of delay-time distribution.
On the other hand, it is also interesting to note that
the mergers of low-mass BH and NS may be more common than NS-NS mergers.
\citet{1998ApJ...506..780B} find that the rate of NS-BH mergers is
20 times larger than that of NS-NS mergers if the initial
mass function (IMF) is supposed to be a Salpeter mass function,
and the mean delay time of a NS-BH binary is $\approx\unit[5]{Gyr}$ \citep[see also][]{2007PhR...442..166N}.
It should be noticed that
the above scenarios are based on primordial binaries, namely,
NS-NS/BH systems that are born as binaries. Alternatively, NS-NS/BH
systems can form dynamically by exchange interactions in globular
clusters during their core collapse. \citet{2006NatPh...2..116G}
estimate that a significant fraction ($\sim30\%$) of NS-NS binaries
may form by this process. The resulting delay time would be dominated
by the timescale of the core-collapse of globular clusters,
which is typically comparable to
the Hubble time ($\tau\approx\unit[6]{Gyr}$ on average; \citealt{2006ApJ...643L..91H}).
The scenario would be more complex, if there are additional populations of SGRBs.
In the context of various assumptions on the delay-time distribution and cosmic star formation
history with the help of the redshift distribution of the observed bursts,
\citet{2006ApJ...650..281N} have
constrained the delay time to be longer than $\unit[4]{Gyr}$,
suggesting that primordial NS-NS progenitors are not favored.
Using a combined
analysis of the luminosity-redshift distribution of SGRBs and the BATSE
logN-logS distribution, \citet{2011ApJ...727..109V} however suggested that
a significant fraction of SGRBs trace the cosmic star formation history with negligible time delays,
implying that there is collapsar contamination in the SGRB population.
By analyzing stellar ages and masses of the host galaxies of 19 SGRBs,
\citet{2010ApJ...725.1202L} found that SGRBs in early- and late-type galaxies seem to have different time delays
with typical delays of $\sim\unit[3]{Gyr}$ and $\sim\unit[0.2]{Gyr}$, respectively.
It should be emphasized that most previous studies
on the delay-time distribution are based on a small number of SGRBs with reliable redshift,
which would seriously limit their ability to obtain a strict constraint on the luminosity function
and the delay-time distribution of SGRBs.

Thanks to the \emph{Swift} satellite, the sample of
SGRBs with measured redshift has significantly increased over
the past seven years.
In this paper, we collected all \emph{Swift} SGRBs with reliable
redshift until 2013 June, which
were selected based on the better criterion of \citet{2013ApJ...764..179B}
to remove possible collapsar contamination.
The luminosity function of SGRBs were then constrained using this sample.
In addition, the prediction
on the progenitor time delays also strongly depends on the star formation rate models.
With different models on the star formation rate, the results on time delays
could change significantly. Especially if the timescale of the delay
is long enough ($\gtrsim\mathrm{Gyr}$),
the star formation at high redshifts, which often differs dramatically in different models, could play a dominate role.
Here, we adopt a self-consistent method in the framework of
hierarchical structure formation to construct the cosmic star formation rate (CSFR).
By using this CSFR with the best-fit luminosity function,
we re-examine the consistency between the observed and expected redshift distribution of SGRBs
under various models for the progenitor delay-time distribution.

This paper is outlined as follows.
In Sect.~2, we elaborate on the details of the star formation models we have used.
In Sect.~3, we describe the method to constrain the luminosity function of SGRBs and to calculate the SGRB rate.
Results are described in Sect.~4, while conclusions are summarized in Sect.~5.
The cosmological parameters used in this paper are $\Omega_{\mathrm{m}}=0.266$,
$\Omega_{\mathrm{\Lambda}}=0.734$, $\Omega_{\mathrm{b}}=0.0449$,
$h=0.71$, and $\sigma_{8}=0.801$.

\section{Model of star formation rate}

We adopt a hierarchical structure formation model from \citet{2010MNRAS.401.1924P},
in which the cosmic star formation rate is derived using the Press-Schechter
(PS) like formalism. To be self-contained,
we give a summary of the most important ingredients of this model in this section.
Following \citet{2010MNRAS.401.1924P},
the equation that governs the total gas density, which includes
the baryon accretion rate, the formation of stars through the transfer
of baryons and the gas ejection by stars, is
determined in a self-consistent way.

The evolution of the total gas density that
controls the star formation history is determined
by the following equation:
\begin{equation}
\dot{\rho}_{g}=-\frac{\mathrm{d}^{2}M_{\star}}{\mathrm{d}V\,\mathrm{d}t}+\frac{\mathrm{d}^{2}M_{ej}}{\mathrm{d}V\,\mathrm{d}t}+a_{b}(t),
\end{equation}
where the first term on the right-hand side is the star formation rate, the
second one is the ejected mass from stars, and the last one represents
the formation of structures through the accretion of baryons from
the intergalactic medium.

The accretion rate of baryons into structures is calculated as follows. In the hierarchical
formation scenario, the distribution of the collapsed objects with
different masses is calculated according to the simple PS formula.
Throughout this paper, we adopt the Sheth-Tormen mass function \citep{1999MNRAS.308..119S},
which is a revised version of the PS mass function,
given by
\begin{eqnarray}
n_{\mathrm{ST}}(M,z)\,\mathrm{d}M & = & A\sqrt{\frac{2a_{1}}{\pi}}\frac{\rho_{\mathrm{m}}}{M}\left[1+\left(\frac{\sigma^{2}}{a_{1}\delta_{c}^{2}}\right)^{p}\right]\frac{\delta_{c}}{\sigma}\nonumber \\
 &  & \exp\left[-\frac{a_{1}\delta_{c}^{2}}{2\sigma^{2}}\right]\frac{\mathrm{d}\ln\sigma^{-1}}{\mathrm{d}M}\,\mathrm{d}M,
\end{eqnarray}
where $A=0.3222$, $a_{1}=0.707$, $p=0.3$, and $\delta_{c}=1.686$.
The parameter $\rho_{\mathrm{m}}$ is the current mean density of the Universe, and
$\sigma$ is the deviation of the linear density field.

The baryon distribution is assumed to directly trace the dark matter distribution,
which means that the density of baryons is just proportional
to the density of dark matter by a factor. Hence, the fraction of
baryons in structures at redshift $z$ is calculated by
\begin{equation}
f_{b}(z)=\frac{\int_{M_{\mathrm{min}}}^{\infty}n_{\mathrm{ST}}(M,z)M\,\mathrm{d}M}{\int_{0}^{\infty}n_{\mathrm{ST}}(M,z)M\,\mathrm{d}M},
\end{equation}
where the threshold mass $M_{\mathrm{min}}$ describes that stars can only form in structures that are suitably dense.
Then the baryon accretion rate $a_{\mathrm{b}}(t)$ that accounts
for the formation of structures can be estimated by
\begin{equation}
a_{\mathrm{b}}(t)=\Omega_{\mathrm{b}}\rho_{\mathrm{c}}\left(\frac{\mathrm{d}t}{\mathrm{d}z}\right)^{-1}\left|\frac{\mathrm{d}f_{\mathrm{b}}(z)}{\mathrm{d}z}\right|.
\end{equation}

The star formation rate is calculated using the Schmidt law \citep{1959ApJ...129..243S},
which gives
\begin{equation}
\frac{\mathrm{d}^{2}M_{\star}}{\mathrm{d}V\,\mathrm{d}t}=\dot{\rho}_{*}(t)=k[\rho_{\mathrm{g}}(t)]^{\alpha},
\end{equation}
where $k$ is a constant, $\rho_{\mathrm{g}}$ is the local gas density,
and $\alpha=1$.

The ejected mass from stars, which is returned to the interstellar
medium, is given by
\begin{equation}
\frac{\mathrm{d}^{2}M_{\mathrm{ej}}}{\mathrm{d}V\,\mathrm{d}t}=\int_{m(t)}^{m_{\mathrm{sup}}}(m-m_{\mathrm{r}})\Phi(m)\dot{\rho}_{*}(t-\tau_{m})\,\mathrm{d}m,
\end{equation}
where $m(t)$ is the mass of a star that has a lifetime of $t$.
The mass of the remnant $m_{\mathrm{r}}$ depends on the
progenitor mass \citep{2010MNRAS.401.1924P}. The stellar IMF $\Phi(m)$ follows the standard \citet{1955ApJ...121..161S}
form, $\Phi(m)=Am^{-2.35}$, with a mass range of $\unit[0.1]{M_{\odot}}<M<\unit[140]{M_{\odot}}$.
The lifetime $\tau_{m}$ of a star with mass $m$ is calculated
using the metallicity-independent fit of \citet{1986FCPh...11....1S}
and \citet{1997ApJ...487..704C}.

Finally, we obtain the function $\rho_{\mathrm{g}}(t)$ at each time $t$,
by combing Eqs.~(4), (5) and (6) with (1). Then the CSFR $\dot{\rho}_{*}(t)$,
according to Eq.~(5), is given by
\begin{equation}
\dot{\rho}_{*}(t)=k\rho_{\mathrm{g}},
\end{equation}
where the constant $k$ is given by the inverse of the timescale
of star formation, namely, $k=1/\tau_{\mathrm{s}}$. The CSFR is
normalized to produce $\dot{\rho}_{*}=\unit[0.016]{M_{\odot}yr^{-1}Mpc^{-3}}$
at $z=0$. We use $\tau_{s}=\unit[2.0]{Gyr}$ as the
timescale for star
formation and $M_{\mathrm{min}}=\unit[10^{8}]{M_{\odot}}$ for the
threshold mass throughout this paper.

\begin{figure}
\resizebox{\hsize}{!}{\includegraphics{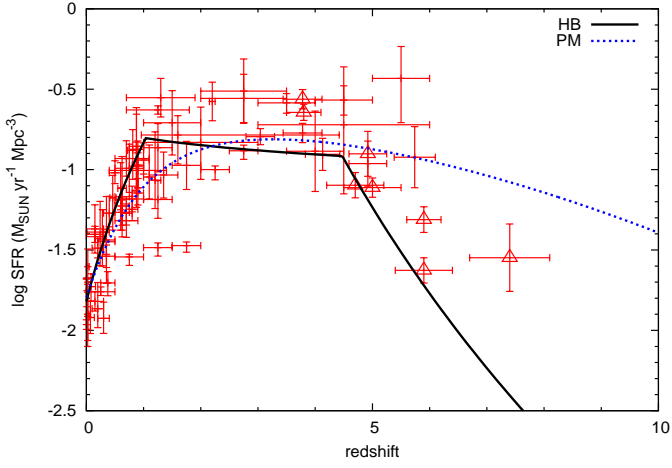}}
\caption{CSFRs as a function of redshift. The dotted line represents the self-consistent model of
\citet{2010MNRAS.401.1924P} (PM), while the solid one shows the best fit of observational
data from \citet{2006ApJ...651..142H} (HB).
The observational data are taken from \citet{2004ApJ...615..209H} (crosses) and
\citet{2008MNRAS.388.1487L} (triangles). }
\label{fig1}
\par\end{figure}

Figure~\ref{fig1} shows the CSFR as a function of redshift. We consider
that the star formation begins at redshift $z_{\mathrm{ini}}=20$.
As can be seen from Fig.~\ref{fig1}, the
fiducial model has an excellent agreement with the observational
CSFR at redshifts $z\lesssim6$. An empirical fit from \citet{2006ApJ...651..142H}
(HB) is also included for comparison. Note that
the self-consistent CSFR remains much flatter than the HB CSFR at redshifts $z\gtrsim4.5$, which
begins to drop exponentially.

\section{Luminosity function and redshift distribution of SGRBs}

\subsection{Sample selection}
\label{sub:sample-selection}

\begin{figure}
\resizebox{\hsize}{!}{\includegraphics{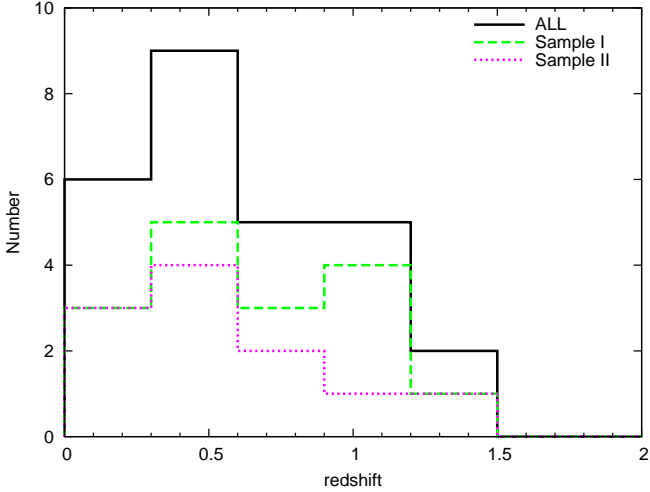}}
\caption{Redshift distributions of \emph{Swift} SGRBs for different samples. The solid histogram represents the full sample of 27 \emph{Swift} SGRBs.
The dashed and dotted histograms are for 16 SGRBs from Sample I ($T_{90}<0.8\,\mathrm{s}$) and 11 SGRBs from Sample II ($T_{90}>0.8\,\mathrm{s}$), respectively.}
\label{fig2}
\par\end{figure}

To investigate the redshift distribution of SGRBs and thus
constrain the delay-time distribution of their progenitors,
we collected all \emph{Swift} GRBs classified as short in the GCN circulars%
\footnote{\href{http://gcn.gsfc.nasa.gov/gcn3_archive.html}{http://gcn.gsfc.nasa.gov/gcn3\_{}archive.html}%
} until 2013 June, which were selected from \citet{2011A&A...529A..97D} and
\citet{2012MNRAS.424.2392K} plus GRB 100206A, GRB 111117A and GRB 130603B
(J. Greiner's web page%
\footnote{\href{http://www.mpe.mpg.de/~jcg/grbgen.html}{http://www.mpe.mpg.de/$\sim$jcg/grbgen.html}%
} and references therein). Only GRBs with well-determined redshift and 15 to 150~keV fluence are included.
According to these criteria,
we obtain a list of 27 GRBs, as shown in Table~\ref{tab:GRB-list}.
It is worth stressing that the classification of a GRB as a short or long burst is complicated, which depends on many
factors such as duration, hardness ratio, spectral lags, etc. Most recently,
\citet{2013ApJ...764..179B} argued that the \emph{Swift} SGRBs that are classically selected according to these factors are
heavily contaminated with collapsars and suggested that a more suitable selection for SGRBs from the \emph{Swift} satellite
be defined by a duration shorter than 0.8~s, which is based on a physically motivated model.
To exclude any possible influence of contaminating collapsars,
we adopt this better criterion and finally obtain a robust sample consisting of 16 SGRBs, which is designated as Sample I.
For comparison, the remaining 11 SGRBs with durations longer than 0.8~s, which have high probability to be collapsars, are considered as Sample II.
The redshift distributions of these two samples are shown in Fig.~\ref{fig2}.

\subsection{Luminosity function of SGRBs}

\begin{figure}
\resizebox{\hsize}{!}{\includegraphics{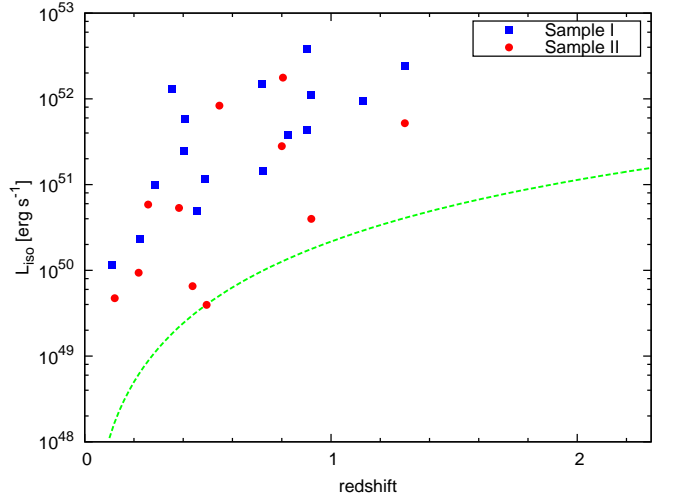}}
\caption{Luminosity-redshift space of \emph{Swift} SGRBs in our samples.
The squares and circles represent the SGRBs from Sample I and Sample II, respectively.
The dashed line represents the flux limit adopted in our calculation,
$F_{\mathrm{lim}}=\unit[5\times10^{-9}]{erg\, s^{-1}\, cm^{-2}}$.}
\label{fig3}
\par\end{figure}

The luminosity of a GRB is computed from the isotropic equivalent
energy ($E_{\mathrm{iso}}$) and the duration of the burst containing
90\% of its total energy ($T_{90}$), using the standard relation:
\begin{equation}
L_{\mathrm{iso}}=\frac{E_{\mathrm{iso}}}{T_{90}/(1+z)},
\end{equation}
where $E_{\mathrm{iso}}$ is calculated in the energy range
$1-10^{4}\,\mathrm{keV}$ in the rest-frame
via a spectral shift procedure described in \citet{2001AJ....121.2879B},
namely,
\begin{equation}
E_{\mathrm{iso}}=\frac{4\pi d_{\mathrm{L}}^{2}}{1+z}Sk(z),
\end{equation}
where $S$ is the fluence in the range of $\unit[15-150]{keV}$ and $k(z)$ is the
$k$-correction defined by
\begin{equation}
k=\frac{\int_{1/(1+z)}^{10^{4}/(1+z)}EN(E)\,\mathrm{d}E}{\int_{15\mathrm{keV}}^{150\mathrm{keV}}EN(E)\,\mathrm{d}E},
\end{equation}
where the observed photon number spectrum $N(E)$ can be well expressed
by a Band function \citep{1993ApJ...413..281B}. The value of $k$
varies from 9.0 to 7.1 as the redshift increases from 0 to 3 with
the peak energy $E_{\mathrm{p}}\sim\unit[490]{keV}$ and low- and
high-energy spectral indices $\alpha=-0.5$ and $\beta=-2.3$, respectively
\citep{2011A&A...530A..21N}. The luminosity-redshift distribution
of our samples is shown in Fig.~\ref{fig3} with the
dashed line indicating the luminosity threshold on the detector's
sensitivity:
\begin{equation}
L_{\mathrm{lim}}(z)=4\pi d_{\mathrm{L}}^{2}k(z)F_{\mathrm{lim}},
\end{equation}
where $d_{\mathrm{L}}(z)$ is the luminosity distance, and the flux
threshold $F_{\mathrm{lim}}$ is taken according to the lowest luminosity
of the sample, which is $F_{\mathrm{lim}}=\unit[5\times10^{-9}]{erg\, s^{-1}\, cm^{-2}}$.

The distribution of $L_{\mathrm{iso}}$ then can be used to
constrain the luminosity function of the SGRBs in our sample (see Fig.~\ref{fig3}).
As there is no theoretical prediction
on the form of the luminosity function of SGRBs, several commonly
used forms are considered, such as the broken power-laws, the Schechter
functions and so on. Among them, only the lognormal function produces a
reliable fitting of the data. Therefore, we use this function
in this work, which reads
\begin{equation}
\Phi(L)=\Phi_{0}\frac{1}{L}\exp\left(\frac{-(\ln L-\ln L_{0})^{2}}{2\sigma^{2}}\right),
\end{equation}
where $L_{0}$ is the mean (peak) value of the luminosity, $\sigma$
is the deviation of the distribution, and $\Phi_{0}$ is a normalization
constant. Note that no evolutionary effects of the luminosity function
are considered in this work.
The best-fit parameters of luminosity function for these two samples are shown in Table~\ref{tab:fits}.
Then we rescale these observed luminosity functions $\Phi(L)$ by the volume
to which the satellite is sensitive. The obtained intrinsic luminosity function
of SGRBs is as follows:
\begin{equation}
\Phi_{\mathrm{intr}}(L)\propto\Phi(L)/d_{\mathrm{L}}^{3}(z_{\mathrm{max}}),
\end{equation}
where $z_{\mathrm{max}}$ is the maximum redshift to which a GRB of
luminosity $L$ can be detected.

\subsection{Modeling the redshift distribution of SGRBs}

The SGRB rate is given by the convolution of the CSFR with the distribution
of the time delays between the star formation and the occurrence
of SGRBs \citep{1992ApJ...389L..45P}:
\begin{equation}
R_{\mathrm{GRB}}(t)\propto\int_{0}^{t(z)}\dot{\rho}_{*}(t-\tau)P(\tau)\,\mathrm{d}\tau,
\end{equation}
where $P(\tau)$ is the probability distribution of the time delays $\tau$.

Because the delay-time distribution $P(\tau)$ is not fully
established theoretically,
we consider the following two simple models that have been widely discussed
in the literature. As mentioned previously, the studies on the rate of type Ia supernovae,
whose progenitors are thought to be double-degenerate
(two white dwarfs) binaries, indicate a power-law delay distribution:
\begin{equation}
P(\tau)\propto1/\tau,
\end{equation}
which also agrees with the observations of six double neutron star binaries \citep{2004MNRAS.350L..61C}
and population synthesis calculations \citep{2006ApJ...648.1110B}.
If the progenitors of SGRBs are dominated by primordial NS-NS binaries,
then this type of delay distribution is our best-guess scenario.
We note that this type of delay distribution was also adopted in the
work of \citet{2005A&A...435..421G}, \citet{2006ApJ...650..281N}, and \citet{2011ApJ...727..109V}.
To investigate other possibilities,
we also consider a lognormal form \citep{2006ApJ...650..281N,2007ApJ...665.1220Z}:
\begin{equation}
P(\tau)=\frac{\exp\{-[\ln(\tau)-\ln(\tau_{*})]^{2}/2\sigma^{2}\}}{\tau\sigma\sqrt{2\pi}},
\end{equation}
with different peak values of $\tau_{*}$ and a narrow ($\sigma=0.3$)
or wide ($\sigma=1.0$) deviation.
This model is useful for gaining some
insight on the constraints of different timescales of delays,
although it is not clear whether it is really related
to the true delay distribution for compact object mergers.

To compare with observations, we calculate
the expected cumulative redshift distribution of the observable SGRBs for different
models:
\begin{equation}
N(<z)=A\int_{0}^{z}\frac{R_{\mathrm{GRB}}(z)}{1+z}\frac{\mathrm{d}V(z)}{\mathrm{d}z}\int_{L_{\mathrm{min}}(z)}\Phi_{\mathrm{intr}}(L)\,\mathrm{d}L,
\end{equation}
where $A$ is a constant and $\mathrm{d}V/\mathrm{d}z$ is the element of the comoving volume per
unit redshift, which is given by
\begin{equation}
\frac{\mathrm{d}V}{\mathrm{d}z}=\frac{4\pi cd_{\mathrm{L}}^{2}}{(1+z)}\left|\frac{\mathrm{d}t}{\mathrm{d}z}\right|.
\end{equation}

\section{Results}

\subsection{Constraints from SGRBs shorter than 0.8~s}

We first consider Sample I, from which SGRBs are selected according to the better
criterion $T_{90}<0.8\,\mathrm{s}$.
Figure~\ref{fig5} shows a comparison between the cumulative
redshift distributions for \emph{Swift} SGRBs from Sample I and the model predictions.
To quantify the consistency between the observed and
expected cumulative redshift distributions of SGRBs for different delay distribution models, an one-sample
Kolmogorov-Smirnov (K-S) test is used.
When assuming that the delay-time distribution is described as a power-law of $\tau^{-1}$,
we find a K-S probability of only 0.04, indicating that
this delay distribution model is disfavored by the observational data.
This is different from the results found for Type Ia supernovae,
which agrees with previous analyses
\citep[e.g.][]{2006ApJ...650..281N,2007ApJ...665.1220Z,2008ApJ...686..408G,2009A&A...498..329G,2011ApJ...727..109V}.
For a narrow lognormal delay distribution, we find that the most likely delay
is \textbf{$\tau_{*}=\unit[5.7]{Gyr}$} ($P\approx0.71$), and its $P>0.05$ interval
is \textbf{$\unit[3.2]{Gyr}<\tau_{*}<\unit[7.7]{Gyr}$}.
While under the assumption of a wide lognormal delay distribution,
the most likely delay is longer than the Hubble time,
and its $P>0.05$ interval is $\tau_{*}>\unit[1.9]{Gyr}$.

Given the observational uncertainty in the CSFR at high redshifts,
we repeat our analysis by using the HB CSFR. As shown in Fig.~\ref{fig1},
the HB CSFR falls exponentially
at redshifts $z\gtrsim4.5$. The results obtained with this CSFR are
similar to those obtained by using the self-consistent CSFR.
For instance, we find that the most likely delay,
when considering a narrow lognormal delay distribution with HB CSFR,
is $\tau_{*}=\unit[4.6]{Gyr}$ ($P\approx0.75$), and its
$P>0.05$ interval is $\unit[2.3]{Gyr}<\tau_{*}<\unit[6.6]{Gyr}$.
It is worth mentioning that the typical delay time we obtained is also smaller
with the smaller star formation rate at high redshifts.
With a better understanding of the distribution of time delays, this implies that
the SGRB rate could also be used to constrain the star formation at high redshifts in the future.

\subsection{Comparison with SGRBs longer than 0.8~s}

As pointed out in Sect.~\ref{sub:sample-selection}
at the suggestion of \citet{2013ApJ...764..179B},
there is a high probability that the physical origin of \emph{Swift} SGRBs from Sample II with
$T_{90}>0.8\,\mathrm{s}$ is different from that of \emph{Swift} SGRBs from Sample I with $T_{90}<0.8\,\mathrm{s}$,
which could lead to
a difference in their delay-time distributions.
To check if these two samples show any significant difference in their delay-time
distributions, we repeat our analysis for Sample II.
The comparison of the redshift distribution of the observed SGRBs from Sample II with the model predictions
is shown in Fig.~\ref{fig6}.
For Sample II,
the power-law ($\tau^{-1}$) delay distribution model is fully consistent with the data ($P\approx0.69$),
in contrast to what is found in Sample I.
Under the assumption of the narrow lognormal delay distribution, the most likely
delay is $\tau_{*}=\unit[4.7]{Gyr}$,
and the $P>0.05$ interval is $\tau_{*}<\unit[11.7]{Gyr}$,
which implies that the progenitors of these SGRBs from Sample II are younger than those from Sample I.
The difference in the delay-time distribution between these two samples could be the result of the
collapsar contamination in the sample of SGRBs with $T_{90}>0.8\,\mathrm{s}$ or,
alternatively, the existence of additional populations.

\section{Conclusions and discussions}

The delay-time distribution of SGRBs is an important
property to single out viable progenitor models.
We presented here a robust sample of 16 SGRBs with reliable redshift and 15 to 150~keV fluence,
which were discovered by the \emph{Swift} satellite until 2013 June.
These SGRBs in our sample are selected according to a better criterion ($T_{90}<0.8\,\mathrm{s}$)
suggested by \citet{2013ApJ...764..179B}, which could
eliminate a very substantial contamination of collapsar GRBs.
Based on this robust sample of \emph{Swift} SGRBs
in the context of various models for the progenitor delay-time distribution and a self-consistent CSFR,
we re-examined whether the model predictions of the redshift distribution of SGRBs are consistent with the observational data.
For this better sample of \emph{Swift} SGRBs with $T_{90}<0.8\,\mathrm{s}$,
we find that the model with a power-law delay distribution of $\tau^{-1}$
shows little consistency with the observational data,
in contrast to the results found for Type Ia supernovae,
which agrees with previous studies
\citep{2006ApJ...650..281N,2007ApJ...665.1220Z,2008ApJ...686..408G,2011ApJ...727..109V}.
We therefore conclude that primordial NS-NS systems are disfavored as the dominant SGRB progenitors.
When considering a model with a narrow lognormal delay distribution, we find that the most likely
delay is $\tau_{*}\sim\unit[5.7]{Gyr}$, and the typical delay is $\tau_{*}>\unit[3.2]{Gyr}$ at a $95\%$ confidence level,
which is relatively shorter than the results of previous analyses that proposed delays longer than $\sim6-7\,\mathrm{Gyr}$
\citep[e.g.][]{2006ApJ...650..281N,2007ApJ...665.1220Z}.
This result implies that the progenitors of SGRBs are dominated by long-lived systems ($\tau\gtrsim\unit[3]{Gyr}$),
which could be understood if dynamically formed NS-NS/BH systems
with an average delay of $\approx\unit[6]{Gyr}$ \citep{2006ApJ...643L..91H}
contribute a significant fraction to the
total number of SGRBs, as proposed by \citet{2008MNRAS.388L...6S}, \citet{2009A&A...498..329G} and \citet{2010ApJ...720..953L}.
Another possible candidate could be primordial NS-BH systems
if these systems do indeed have an average delay of $\approx\unit[5]{Gyr}$
\citep[see][]{1998ApJ...506..780B,2007PhR...442..166N}.

We also tested whether there is any difference between \emph{Swift} SGRBs with $T_{90}<0.8\,\mathrm{s}$ and those with $T_{90}>0.8\,\mathrm{s}$,
which is expected if the \emph{Swift} SGRBs with $T_{90}>0.8\,\mathrm{s}$ are heavily contaminated by collapsars.
We find that \emph{Swift} SGRBs from the $T_{90}>0.8\,\mathrm{s}$ sample have shorter delays than those with $T_{90}<0.8\,\mathrm{s}$,
which could be interpreted as the contamination by collapsars, as suggested by \citet{2011ApJ...727..109V} and \citet{2013ApJ...764..179B}.
However, the possibility of the existence of additional populations cannot be excluded.
We caution that a more detailed comparison of the host galaxies of \emph{Swift} SGRBs with $T_{90}>0.8\,\mathrm{s}$ to those of LGRBs is needed before
any firm conclusion can be drawn, as also suggested by \citet{2010ApJ...725.1202L}.

Although the sample of \emph{Swift} SGRBs with reliable redshift has significantly
expanded in the past seven years, their total number is still small, which
prohibits a strict constraint on their luminosity function
and delay-time distribution.
The importance of the contribution of dynamically formed NS-NS systems and NS-BH systems
would only be severely constrained by the detection of more high-redshift
($z>1$) SGRBs. Detailed observations of the host galaxies of individual SGRB
are also essential to provide a better description of the distribution of time delays.
Because these merging binary systems are also
one of the most powerful sources of GWs, it is expected
that the detection of GW signals from these sources
would be helpful for validating different theoretical models.
In particular, it would be relatively easy for a GW detector to distinguish between these two type of sources and then to constrain their relative contribution to the occurrence of SGRBs, models, since NS-BH mergers emit more powerful and lower frequency GWs than NS-NS mergers.  

\begin{acknowledgements}
We thank the anonymous referee for her/his useful suggestions, which
have significantly improved this paper.
We also thank J. Greiner for his online GRB Table.
This work is partially supported by
National Basic Research Program of China (2009CB824800, 2012CB821800),
the National Natural Science Foundation (11073020, 11133005, 11233003),
and the Fundamental Research Funds for the Central Universities (WK2030220004).
\end{acknowledgements}


\clearpage{}

\begin{figure}
\resizebox{\hsize}{!}{\includegraphics{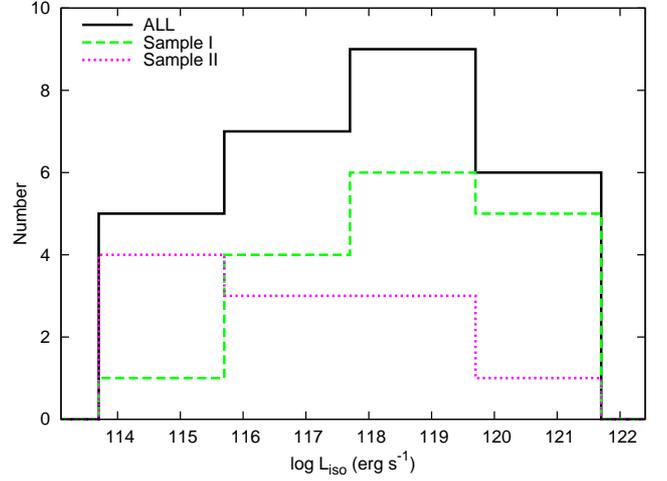}}
\caption{Number distributions of the luminosities of \emph{Swift} SGRBs in our samples.}
\label{fig4}
\par\end{figure}

\begin{figure}
\resizebox{\hsize}{!}{\includegraphics{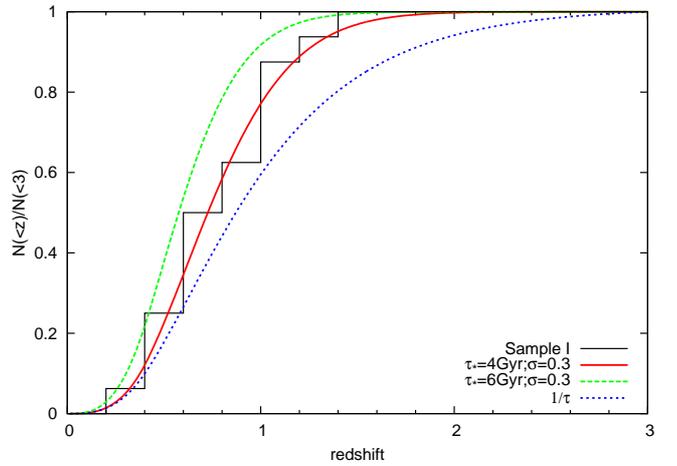}}
\caption{Comparison between the observed and expected cumulative
distributions of SGRBs with several representative delay-time distributions for Sample I.}
\label{fig5}
\par\end{figure}

\begin{figure}
\resizebox{\hsize}{!}{\includegraphics{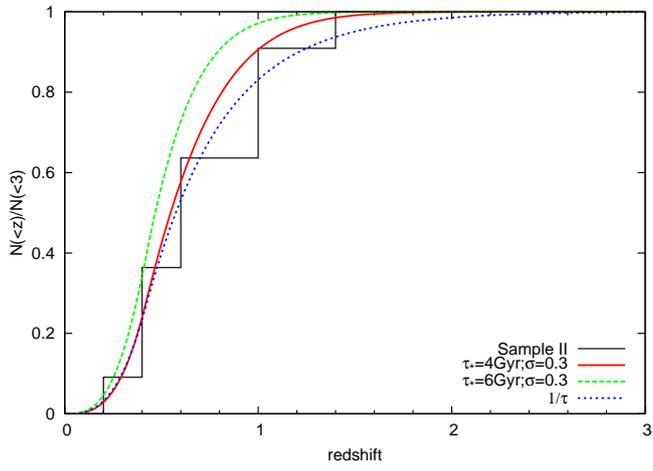}}
\caption{Same as Fig.~\ref{fig5}, but for \emph{Swift} SGRBs from Sample II.}
\label{fig6}
\par\end{figure}

\clearpage{}

\begin{table}
\caption{Best-fit parameters of SGRB luminosity function.}
\label{tab:fits}
\centering
\begin{tabular}{ccc}
\hline\hline
Sample & $L_{0}$ & $\sigma$\\
 & ($10^{51}\,\mathrm{erg\, s^{-1}}$) & \\
\hline
I & 3.16 & 1.68\\
II & 0.59 & 2.19\\
\hline
\end{tabular}
\end{table}

\begin{table}
\caption{List of \emph{Swift} SGRBs in our sample.}
\label{tab:GRB-list}
\centering
\begin{tabular}{ccccc}
\hline\hline
GRB & Redshift & Duration  & Fluence  & Ref\\
 & &(s)&($\unit[10^{-7}]{erg\, cm^{-2}}$)&\\
\hline
050509b & 0.225 & 0.05 & $0.09\pm0.02$ & 1,2\\
050724  & 0.2576  & 3.0 & $9.98\pm1.20$ & 1,2\\
050813  & 0.722  & 0.6 & $0.44\pm0.11$ & 1,2\\
051210  & 1.3 & 1.3 & $0.85\pm0.14$ & 2\\
051221a  & 0.547  & 1.4 & $11.50\pm0.35$ & 1,2\\
060502b  & 0.287  & 0.09 & $0.40\pm0.05$ & 1,2\\
060801  & 1.131  & 0.5 & $0.80\pm0.10$ & 1,2\\
061006  & 0.4377  & 130 & $14.20\pm1.42$ & 1,2\\
061201  & 0.111  & 0.8 & $3.34\pm0.27$ & 1,2\\
061217  & 0.827  & 0.3 & $0.42\pm0.07$ & 1,2\\
070429b  & 0.904  & 0.5 & $0.63\pm0.10$ & 1\\
070714b  & 0.92  & 64.0 & $7.20\pm0.90$ & 1,2\\
070724a  & 0.457  & 0.4 & $0.30\pm0.07$ & 1,2\\
070729  & 0.8  & 0.9 & $1.00\pm0.20$ & 2\\
070809  & 0.2187  & 1.3 & $1.00\pm0.10$ & 2\\
070810b  & 0.49  & 0.08 & $0.12\pm0.03$ & 1,2\\
071227  & 0.383  & 1.8 & $2.20\pm0.30$ & 1,2\\
080123  & 0.495  & 115.0 & $5.70\pm1.70$ & 2\\
080905  & 0.1218  & 1.0 & $1.40\pm0.20$ & 2\\
090510  & 0.903  & 0.3 & $3.40\pm0.40$ & 1,2\\
090515  & 0.403  & 0.036 & $0.20\pm0.03$ & 2\\
100117a  & 0.92  & 0.3 & $0.93\pm0.13$ & 2\\
100206a  & 0.4068  & 0.12 & $1.40\pm0.20$ & 3\\
100816a  & 0.8049  & 2.9 & $20.00\pm1.00$ & 2\\
101219a  & 0.718  & 0.6 & $4.60\pm0.30$ & 2\\
111117a & 1.3 & 0.47 & $1.40\pm0.18$ & 3\\
130603b & 0.3564 & 0.18 & $6.30\pm0.30$ & 3\\
\hline
\end{tabular}

\tablebib{(1) \citet{2011A&A...529A..97D}; (2) \citet{2012MNRAS.424.2392K};
(3) Greiner's GRB page}
\end{table}

\end{document}